\documentstyle[twoside,fleqn,espcrc2]{article}

\title{\vbox{\hbox{\rightline{\rm\small EFI-97-25}}
\hbox{\rightline{\rm\small hep-th/9706039}}
\hbox{Spin Dependence of D0-brane Interactions}}}

\author{Jeffrey A. Harvey \address{Enrico Fermi Institute and Department
        of Physics \\
        University of Chicago, 5640 Ellis Ave., Chicago, IL 60637}%
        \thanks{Supported in part by  NSF Grant No.~PHY 9600697}}

\begin{document}

\begin{abstract}
The long-range, spin-dependent forces between D0-branes are
related to long-range fundamental string interactions using duality.
These interactions can then be computed by taking the long distance
non-relativistic expansion of string four-point amplitudes.
The results are in accord with the general constraints
of Matrix Theory.
\end{abstract}

\maketitle

\section{Introduction}

It is commonly stated that there is a vanishing static force between
BPS states in supersymmetric theories and this is  reflected in the
fact that there often exist exact multiparticle solutions to the classical BPS
equations for supersymmetric configurations. This simple picture is
however modified by quantum and relativistic effects. Quantum effects
associated with fermion zero modes give soliton multiplets with spin which
lie in representations of the spacetime supersymmetry algebra. Spin
dependent static forces between BPS states are then formally of the same
order in an expansion in powers of $1/c$ as velocity dependent interaction
terms which are independent of spin, a fact which should
be familiar from the study of relativistic corrections to the
hydrogen atom spectrum where $p^4$ corrections to the Hamiltonian
and spin-orbit interactions arise at the same order in $1/c$.
Thus we must  include these spin-dependent interactions to have
a consistent supersymmetric description of the long-range interactions
between BPS states.

An example of this phenomenon occurs in the long-range interactions
between magnetic monopoles in $N=4$ gauge theory. The supersymmetric
quantum mechanics on the two monopole moduli space contains a four
fermion term proportional to the curvature tensor on the moduli space.
Upon quantizing the fermion fields this interaction gives a static
spin-dependent interaction between two monopoles. The existence of this
long-range interaction is crucial for the existence of BPS bound states
in this system as can be seen from the analysis in \cite{sendual}.

In this talk I will discuss the long-range spin dependent interactions
between D0-branes in type IIA string theory.

\section{D0-branes and duality}

We first recall a few basic facts about D0-branes. These are BPS states
which fall into a 256 dimensional short representation of the $N=(1,1)$
supersymmetry algebra in $D=10$. This representation consists of states
transforming as $128 \oplus 84 \oplus 44$ under the $SO(9)$ rotation group.
The D0-branes can be viewed as the Kaluza-Klein momentum modes of the
Rarita-Schwinger field ($\psi_M$), third rank antisymmetric tensor field
($A_{MNP}$), and graviton field ($g_{MN}$) of $D=11$ supergravity
compactified on $S^1$ \cite{townsend,wittendyn}.

The leading spin independent static force between D0-branes vanishes
as shown in \cite{dnotes}.
The leading non-zero long-range interaction between D0-branes
goes as $v^4/r^7$ with
$v$ the D0-brane velocity and $r$ their separation \cite{dkps}. The
fact that Matrix theory reproduces this interaction through a one-loop
calculation in quantum mechanics was one of the first pieces of evidence
for the correctness of the proposal of \cite{bfss}. The $v^4/r^7$
interaction is universal, that is independent of the spin state of
the D0-brane. There are a number of approaches to computing the spin
dependent long-range interactions. At large distances only exchanges
of massless states contribute so one needs only to compute the t-channel
diagrams from exchange of the massless fields of IIA supergravity. The
various normalizations that enter into such a computation are most easily
organized by performing the computation directly in string theory. This
could be done by generalizing the computation of \cite{dnotes} to include
the effects of fermion boundary states. Here we will take a more indirect
route using string duality. The spin-dependent interactions should also
follow from Matrix theory by generalizing the computation of \cite{dkps}
to include the terms which are required to supersymmetrize the $v^4/r^7$
interaction.

To compute the spin dependent interactions we first compactify the IIA
theory on a $S^1$ of radius $R_9$ with coordinate $x_9$. A T-duality
transformation along $x_9$ turns the D0-brane into a D1-brane wrapped
on the $S^1$ in IIB theory. An S-duality transformation then turns
the D1-brane into a fundamental IIB string wrapped on the $S^1$. Since
the long-range forces between BPS states are dictated by their coupling
to massless states and these couplings are protected by supersymmetry, the
long-range interaction between D0-branes can be computed by computing
the long-range interactions between wound IIB fundamental strings.
One disadvantage of this approach is that massive states in $D=9$
are classified by $SO(8)$ so that the answer will be expressed in
terms of $SO(8)$ invariants rather than $SO(9)$ invariants.

\section{Kinematics}

For IIB string on $S^1$ half-saturated BPS states satisfy
$N_L=N_R=0$ (in Green-Schwarz formalism)
and hence $mn=0$ with $m$ the momentum and $n$ the
winding on $S^1$. We consider $m=0$ and $n$ arbitrary. Under the duality
described above $n$ labels the D0-brane charge.
The $SO(9)$ spin states for D0-branes
decompose under $SO(9) \rightarrow SO(8)$ as
\begin{eqnarray}
{\bf 44} & \rightarrow & ({\bf 1  + 35_v})_{NSNS} +
                                           ({\bf 8_v})_{RR} \\
                    {\bf 84} & \rightarrow & ({\bf 28})_{NSNS} +
                                           ({\bf 56_v})_{RR} \\
             {\bf 128} & \rightarrow & ({\bf 8_c + 8_s + 56_c +
             56_s})_{NSR+RNS}
\end{eqnarray}
where the subscripts indicate which sector (RR, NSNS, RNS, NSR) of IIB
string theory the states come from.

We want to compute four-point string amplitudes for the scattering of
 such states. These states are labelled by the left and right-moving
spinors or polarization vectors to be discussed momentarily and by their
ten-dimensional momenta

\begin{eqnarray}
p_1^{R,L} & = &({1 \over 2} \hat p_1,\pm nR) \\
                     p_2^{R,L} & = & ({1 \over 2} \hat p_2,\pm nR) \\
                     p_3^{R,L} & = & ({1 \over 2} \hat p_3,\mp nR) \\
                     p_4^{R,L} & = & ({1 \over 2} \hat p_4,\mp nR)
\end{eqnarray}
Here the hatted quantities are nine-dimensional momenta which
in the center of mass frame are given by
\begin{eqnarray}
\hat p_1 & = & E(1, \vec v), \qquad \quad
\hat p_2 = E(1,- \vec v) \\
 \hat p_3 & = &  -E(1, \vec w), \qquad \hat p_4 = -E(1,- \vec w)
\end{eqnarray}
with $|\vec v| = | \vec w| = v$, $\vec v \cdot \vec w = v^2 \cos \theta$,
and $E^2(1-v^2) = M^2 \equiv 4 n^2 R^2$. It will also be useful to
define the eight-momentum transfer $\vec q = M(\vec v + \vec w)$
and the vector $\vec k = M(\vec v - \vec w)$ which as $\vec q$ goes to
zero reduces to $2M \vec v$ with $\pm \vec v$ the velocity of the
incoming D0-branes.

Note that the ten-dimensional Mandelstam invariants
are the same for left and right-movers
since there are no directions with both momentum and winding.

We also define nine-dimensional Mandelstam invariants
\begin{eqnarray}
 \hat s & \equiv & - (\hat p_1 + \hat p_2)^2 \\
                       \hat t & \equiv & - (\hat p_2 + \hat p_3)^2 \\
                       \hat u & \equiv & - (\hat p_1 + \hat p_3)^2
\end{eqnarray}

\subsection{Polarization vectors and spinors}

We will also need some details about  polarization
vectors and spinors.  With the above kinematics we can
consider a t-channel process in which
particle $1$ exchanges massless fields with particle $2$
and the particles labelled $4$ and $3$ are the final states of
particles $1$ and $2$ respectively. To compute the long-range
force we are interested in a configuration where the spins of
particles $1$ and $4$ are the same in their rest frame
as are those of particles
$2$ and $3$. Because we need to compute at non-zero momentum
transfer, the spins of particles $1$ and $4$ (and $2$ and $3$)
will not be the same, rather they will differ by the fact
that we need to boost their identical rest frame spins by
different velocities.
Thus with the previous center of mass kinematics the polarization
vectors labelling the bosonic components of the left or right-moving
states will be
boosts of a
polarization vector $\vec \zeta_1$ by $\vec v$ and $- \vec w$ respectively
and similarly for particles 2 and 3. To the order in velocity required
here this gives
\begin{eqnarray}
\lefteqn{\hat p_1  =  M(1 + v^2/2, \vec v), \quad
\zeta_1 = (\vec \zeta_1 \cdot \vec v, \vec \zeta_1)} \\
\lefteqn{\hat p_2  = M(1+ v^2/2, - \vec v), \quad
\zeta_2 = (-\vec \zeta_2 \cdot \vec v, \vec \zeta_2)} \\
\lefteqn{\hat p_3   = - M(1 + w^2/2, \vec w), \quad
\zeta_3 = (\vec \zeta_2 \cdot \vec w, \vec \zeta_2)} \\
\lefteqn{\hat p_4   = - M(1 + w^2/2, - \vec w), \quad
\zeta_4 = (- \vec \zeta_1 \cdot \vec w, \vec \zeta_1)}
\end{eqnarray}

Similar remarks apply to the spinors which label these
states, that is we again take them to be equal for particles
$1$ and $4$ and for $2$ and $3$ in the rest frame and then
boost these spinors by $\pm v$, $\pm w$ to obtain the
correct center of mass frame spinors. Our
conventions for ten-dimensional gamma matrices are as follows. Define
\begin{equation}
\bar \gamma^i = i \left( \begin{array}{rr}
 0 & \gamma^i \\
 \gamma^i & 0 \\
\end{array} \right)
\end{equation}
where the $\gamma^i$ are as in \cite{gsw}.
 We take
\begin{equation}
\bar \gamma^9 = i \left( \begin{array}{rr}
 1 & 0 \\
0 & -1 \\
\end{array} \right)
\end{equation}
The 32 by 32 $SO(9,1)$ gamma matrices are then
\begin{equation}
\Gamma^0 = i \left( \begin{array}{rr}
 0 & 1 \\
 -1 & 0 \\  \end{array} \right), \qquad
\Gamma^m = \left( \begin{array}{rr}
 0 & \bar \gamma^m \\ \bar \gamma^m & 0 \\
\end{array} \right)
\end{equation}
with $m = 1, \ldots 9$. Note that this representation is purely imaginary.
We also define
\begin{equation}
\Gamma^{11} = - \Gamma^0 \Gamma^1 \cdots \Gamma^9 =
{diag}(1,-1)
\end{equation}
In IIB string theory we start with both left and right spinors tramsforming
as the $16_+$ of $SO(9,1)$, that is with $\Gamma^{11}$ eigenvalue $+1$.

The Dirac equation is
\begin{equation}
\Gamma^\mu p^{L,R}_\mu u^{L,R}(p) =0
\end{equation}
where it is to be understood that $u$ transforms as $16_+$ and hence
can be written as the transpose of $(\lambda^{L,R},0)$ with $\lambda^{L,R}$
a 16 component spinor. In the rest frame we have
$p^{R,L}=(M/2,\vec 0, \pm R)$ so that
\begin{equation}
(M/2)(\Gamma^0 \mp \Gamma^9)u^{R,L} = 0
\end{equation}
which using the given basis for the gamma matrices is equivalent
to the equation
\begin{equation}
-i \bar \gamma^9 \lambda^{R,L} = \mp \lambda^{R,L}
\end{equation}
Thus if we define the $8_s$ of the $SO(8)$ little group of a massive
particle to be states with $-i \bar \gamma^9$ eigenvalue $+1$ then this shows
that $\lambda^R \sim 8_c$ and $\lambda^L \sim 8_s$.
Note that we could also have used momentum states in the IIA string
and would have found the same representations. Thus we conclude
that the rest frame spinors are given by
\begin{equation}
\lambda^L = \sqrt{M/2} \left(\begin{array}{c}
\xi_a^L\\
0 \\
\end{array} \right), \quad \lambda^R = \sqrt{M/2} \left(\begin{array}{c}
0 \\
\xi_{\dot a}^R \\
\end{array} \right )
\end{equation}

Boosting then gives
the
left-handed spinors to order velocity
squared:
\begin{equation}
 \lambda_1^L  = \sqrt{M/2} \left( \begin{array}{c}
(1 + v^2/4) \xi_1^L \\
-( \vec v \cdot \vec \gamma / 2) \xi_1^L \\
\end{array} \right )
\end{equation}
\begin{equation}
  \lambda_2^L  = \sqrt{M/2} \left( \begin{array}{c}
(1 + v^2/4) \xi_2^L \\
( \vec v \cdot \vec \gamma /2 ) \xi_2^L \\
\end{array} \right)
\end{equation}
\begin{equation}
 \lambda_3^L  = \sqrt{M/2} \left(\begin{array}{c}
(1 + w^2/4) \xi_2^L \\
-( \vec w \cdot \vec \gamma / 2) \xi_2^L \\
\end{array} \right)
\end{equation}
\begin{equation}
 \lambda_4^L  = \sqrt{M/2} \left( \begin{array}{c}
(1 + w^2/4) \xi_1^L \\
( \vec w \cdot \vec \gamma / 2) \xi_1^L \\
\end{array} \right)
\end{equation}
with similar formulae for the right-moving components.

\section{Four point string amplitude}

The four point amplitude is calculated by sewing together two open
string four point amplitudes  using the
formalism of \cite{klt}. This gives
\begin{equation}
A(1,2,3,4) = \kappa^2 \sin(\pi \hat t/8)
A_4^{L} \times A_4^{R}
\end{equation}
where $A_4^{L,R}$ are open string four point functions.
In terms of nine-dimensional kinematics we have
\begin{equation}
A_4^{L}  =  {\Gamma(-\hat s/8+M^2/2)
\Gamma(-\hat t/8) \over \Gamma(1-\hat t/8 -\hat s/8 +M^2/2)} K^L(1,2,3,4)
\end{equation}
and
\begin{equation}
A_4^{R}  =  {\Gamma(- \hat t/8) \Gamma(- \hat u/8)
\over \Gamma(1-\hat t/8 -\hat u/8) } K^{R}(1,2,3,4)
\end{equation}
where $K^{L,R}(1,2,3,4)$ are  kinematical factors depending on
the momenta, polarization vectors, and  spinors depending on
whether the left or right-moving amplitude involves scattering
of four vector states, four fermion states, or two vectors and
two fermions. We will denote these as $K^L(\zeta_{1L},\zeta_{2L},\zeta_{3L},
\zeta_{4L})$, $^LK(u_{1L},u_{2L},u_{3L},u_{4L})$ and
$K^L(u_{1L},\zeta_{2L},\zeta_{3L},u_{4L})$ respectively and similarly for the
right-moving factor.

Expanding the sine and gamma functions  in the non-relativistic
limit gives
\begin{eqnarray}
\lefteqn{ A(1,2,3,4)  \sim } \\
& & {\kappa^2 \pi 8^3 \over \hat u \hat t (\hat s - 4M^2)}
K^L(1,2,3,4)K^R(1,2,3,4)
\end{eqnarray}
In order to extract the long-range interactions we need to
extract the $\hat t$ channel poles in this expression and
then Fourier transform  with respect to $\vec q$.
As a result, any terms in $K^{L,R}$ which are proportional
to $\hat t$ will not contribute to the $\hat t$ channel pole
and can be dropped for the purpose of extracting the
long-range interactions (they may contribute to contact interactions).

As an example consider the interaction of two states arising
from the $(NS)^2$ sector. In this case we need the non-relativstic
expansion of $K^L(\zeta_{1L},\zeta_{2L},\zeta_{3L},\zeta_{4L})
K^R(\zeta_{1R},\zeta_{2R},\zeta_{3R},\zeta_{4R})$. Using the previous
expansions for the polarization tensors and momenta and dropping
terms proportional to $\hat t$ we find
\begin{eqnarray}
\lefteqn{K^L(\zeta_{1L},\zeta_{2L},\zeta_{3L},\zeta_{4L})  =
{\hat u \over 64} \times } \\
 & & \left[ \vec k^2 - 2(\vec q \cdot \vec \zeta_{1L})^2 -
2(\vec q \cdot \vec \zeta_{2L})^2 \right]
\end{eqnarray}

For the scattering of two $(NS)^2$ states we then have for the
t-channel pole
\begin{equation}
A_4 = - {\pi \kappa^2 \over 8 \vec q^2}
\left[ \vec k^2 - 2(\vec q \cdot \vec \zeta_{1})^2 -
2(\vec q \cdot \vec \zeta_{2})^2 \right]^2
\end{equation}
where the square on the term in square brackets indicates
the product of two terms with left and right-moving
polarization vectors respectively.

Fourier transforming this amplitude with respect to $\vec q$, then
gives the  potential
\begin{eqnarray}
\lefteqn{V \sim } \\
& & \left[ 4 M^2 \vec v^2 + 2(\vec \nabla \cdot \vec
\zeta_{1})^2 +
2(\vec \nabla \cdot \vec \zeta_{2})^2 \right]^2 {1 \over r^6}
\end{eqnarray}

This exhibits the spin-independent $v^4$ interaction, here going like
$1/r^6$ instead of $1/r^7$ because we have dimensionally reduced to
$D=9$. In addition we see that there is a  spin-dependent term
proportional to $v^2$ which goes like $1/r^8$ ($1/r^9$ in $D=10$)
and a static spin-dependent term going like $1/r^{10}$ ($1/r^{11}$ in $D=10$).
Note that since this expression is symmetric in $\zeta_L$ and $\zeta_R$
the spin-dependent terms vanish for  $(NS)^2$ states in
the ${\bf 28}$ of $SO(8)$ since this occurs in the
antisymmetric tensor product of ${\bf 8_v} \times
{\bf 8_v}$.

As a second example consider the interaction of two bosons which
arise from the $(R)^2$ sector. For these states we need the
non-relativistic expansion of
\begin{equation}
K^L(u_{1L},u_{2L},u_{3L},u_{4L}) K^R(u_{1R},u_{2R},u_{3R},u_{4R})
\end{equation}
with
\begin{equation}
K(u_1,u_2,u_3,u_4)  = \left[ (-\hat s/8 +M^2/2)
\bar u_2 \Gamma^\mu u_3 \bar u_1 \Gamma_\mu u_4 \right]
\end{equation}
where we have again dropped terms proportional to $\hat t$.

Using the previous expressions for the spinor fields one finds
after some algebra that
\begin{eqnarray}
\lefteqn{K^L(u_{1L},u_{2L},u_{3L},u_{4L})  = {\hat s - 4 M^2 \over 8}
\times } \\
\lefteqn{ \left[
 {\vec k^2 \over 8} +{5 \over 16} k_i q_j (R^{ij}_{1L} + R^{ij}_{2L})
- {1 \over 16} q_j q_k R^{ij}_{1L} R^{ik}_{2L} \right] }
\end{eqnarray}
and similarly for right-movers where $R_{1L}^{ij} = \xi_{1L}^{\dagger}
\gamma^{ij} \xi_{1L}/4$. Combining left and right-moving parts
then gives  for the four point amplitude
\begin{eqnarray}
\lefteqn{A_4 = - {\pi \kappa^2 \over 8 \vec q^2} \times } \\
\lefteqn{ \left[
 {\vec k^2 } +{5 \over 2} k_i q_j (R^{ij}_{1} + R^{ij}_{2})
- {1 \over 2} q_j q_k R^{ij}_{1} R^{ik}_{2} \right]^2 }
\end{eqnarray}

We again see the universal spin-independent $v^4$ term as in the
expression for scattering of $(NS)^2$ states and additional spin
dependent terms involving $v^3$, $v^2$, $v$, and a static term.
Extrapolating to $D=10$ the Fourier transform of $A_4$
has the schematic form
\begin{equation}
V \sim {v^4 \over r^7} + {v^3 R \over r^8} + {v^2 R^2 \over r^9}
+ {v R^3 \over r^{10}} + {R^4 \over r^{11}}
\end{equation}
with $R$ a tensor bilinear in spinor fields.

Scattering of fermionic D0-branes in the R-NS sector can also
be computed in an analogous manner. We consider scattering
of states in the $NS-R$ sector with states in the $NS-R$ sector.
The four point amplitude then factorizes as in eqn. (34) with
$K_L$ given by eqn. (35) and $K_R$ by eqn. (41). The four point
amplitude expanded in velocity as before is then
\begin{eqnarray}
\lefteqn{A_4 =  - {\pi \kappa^2 \over 8 \vec q^2} \times } \\
\lefteqn{\left[ \vec k^2 - 2(\vec q \cdot \vec \zeta_{1})^2 -
2(\vec q \cdot \vec \zeta_{2})^2 \right] \times } \\
\lefteqn{ \left[
 {\vec k^2 } +{5 \over 2} k_i q_j (R^{ij}_{1} + R^{ij}_{2})
- {1 \over 2} q_j q_k R^{ij}_{1} R^{ik}_{2} \right] }
\end{eqnarray}
Fourier transforming with respect to $\vec q$ then gives the
spin dependent potential between fermionic D0-branes.

\section{Comparison with Matrix theory}

These interactions should also be computable from the Matrix model
proposed in \cite{bfss} by generalizing the one-loop computation of
\cite{dkps,beckers} to include the fermion terms which are related to
the $v^4/r^7$ term by supersymmetry. Such a computation has not yet
appeared in the literature, so here we will be content to compare these
results with the structure of the matrix model which can be determined
from simple scaling arguments.

If we rescale fields
so that the matrix model action takes the form
(with $\phi$ the boson fields, $\psi$ the fermion fields and $\partial$
denoting time derivative)
\begin{equation}
 S = {1 \over g_s} \int dt \left( (\partial \phi)^2 + \phi^4 + \psi
\partial \psi + \psi^2 \phi \right)
\end{equation}
then we can do dimensional analysis by assigning the following
dimensions to the fields, couplings, and derivatives:
\begin{equation}
[g_s] = -3, \quad [\phi] = -1, \quad [\partial] = -1,
\quad [\psi] = -3/2
\end{equation}
The loop expansion for the effective action then has the form
\begin{equation}
S = {g_s}^{-1} S_0 + S_1 + {g_s} S_2 + \cdots
\end{equation}
and the Lagrangian density ${\cal L}_i$  has dimension $-4+3i$.

We can also organize terms by the parameter $N= n_{\partial} + 2 n_f$
with $n_{\partial}$ the number of time derivatives and $n_f$ the number
of fermion fields. The supersymmetry transformations preserve $N$ so
terms paired by supersymmetry must all have the same value of $N$.
The $v^4/r^7$ term occurs at one-loop and thus has dimension $-1$.
It also cleary has $N=4$. If all the terms we have computed have
dimension $-1$ and $N=4$ then it is plausible that they all arise
from supersymmetrization of the $v^4/r^7$ term. To check this note
that in the matrix model both the spinor bilinear $R$ and any polarization
tensors $\zeta$ have to arise as expectation values of operators
which are bilinear in the fermion fields  $\psi$. The potential
terms we have computed when extrapolated to $D=10$ then all have
dimension $-1$ and $N=4$ and match
on to terms in the one-loop matrix model effective action
of the schematic form
\begin{equation}
{\cal L}_1 \sim {v^4 \over r^7} + {v^3 \psi^2 \over r^8}
+ {v^2 \psi^4 \over r^9} + {v \psi^6 \over r^{10} } + {\psi^8 \over r^{11}}
\end{equation}

\section{Acknowledgements}

I would like to thank
H. Awata, S. Chaudhuri, M. Li, D. Minic, G. Moore,  E. Novak,
J. Polchinski, and S. Shenker for discussions.  I am particularly grateful
to C. Bachas and M. Green for discussions of their work on similar
topics. I would also like to thank the organizers of Strings '97 for
the invitation to present these results.

\end{document}